\documentclass[
 reprint,
superscriptaddress,
prl,
 amsmath,amssymb,
 aps,
]{revtex4-2}

\usepackage{graphicx}
\usepackage{dcolumn}
\usepackage{bm}
\usepackage{hyperref}
\usepackage{xcolor}
\usepackage{siunitx}

\begin{document}

\preprint{APS/123-QED}

\title{Information Thermodynamics in a Quantum Dot Szilard Engine - Experimentally Investigating Fluctuation Theorems and Thermodynamic Uncertainty Relations}
	\author{David Barker}
	\affiliation{NanoLund and Solid State Physics, Lund University, Box 118, 22100 Lund, Sweden}
	\author{Sebastian Lehmann}
	\affiliation{NanoLund and Solid State Physics, Lund University, Box 118, 22100 Lund, Sweden}
	\author{Kimberly A. Dick}
	\affiliation{Centre for Analysis and Synthesis, Lund University, Box 124, 22100 Lund, Sweden}
	\author{Peter Samuelsson}
	\affiliation{Physics Department and NanoLund, Lund University, Box 118, 22100 Lund, Sweden}
	\author{Ville F. Maisi}
	\affiliation{NanoLund and Solid State Physics, Lund University, Box 118, 22100 Lund, Sweden}
	\author{Patrick P. Potts}
	\affiliation{Department of Physics and Swiss Nanoscience Institute, University of Basel, Klingelbergstrasse 82, 4056 Basel, Switzerland}

\date{\today}

\begin{abstract}
In Szilard's engine, measurement and feedback allows to extract work from an equilibrium environment, a process otherwise forbidden by the laws of thermodynamics. Recent theoretical developments have established fluctuation theorems and thermodynamic uncertainty relations that constrain the fluctuations in Szilard's engine. These relations rely on auxiliary experimental protocols known as backward experiments. Here, we experimentally investigate the thermodynamics of Szilard's engine by implementing two distinct types of backward experiments. We verify and compare the corresponding fluctuation theorems and thermodynamic uncertainty relations associated with each protocol. Our results reveal that the entropy production inferable from measurement may serve as a more relevant quantifier of information than the widely used mutual information. 
\end{abstract}

\maketitle

\textit{Introduction}--Feedback control \cite{wiseman:2010,zhang:2017} is an important ingredient for emerging quantum technologies with applications ranging from state preparation \cite{magrini:2021} to quantum error correction \cite{krinner:2022}. A thermodynamic description of feedback-controlled systems provides a deeper understanding of their capabilities and limitations, paving the way for novel applications. 
Famously, already the thought experiment by James C. Maxwell \cite{maxwell:1871,MaruyamaMD2009,rex:2017}, which illustrates that measurement and feedback enables processes that are otherwise prohibited by the laws of thermodynamics, established the deep link between thermodynamics and information. A paradigmatic example of such a process is provided by Szilard's engine, which uses information about a single particle to extract work \cite{szilard:1929,parrondo_thermodynamics_2015}. 
During the last decades Szilard's engine has been implemented in a variety of systems including Brownian particles \cite{toyabe_experimental_2010,roldan:2014}, metallic islands \cite{koski:2014,koski_experimental_2014}, and semiconducting quantum dots~\cite{BarkerFDR2022,aggarwal:2024}.

Feedback control exploits the stochastic nature of a system, with different measurement outcomes resulting in different feedback protocols. In the last decades, the thermodynamics of stochastic systems has been extensively studied and the fluctuation theorem (FT) was established as the generalization of the second law of thermodynamics for stochastic systems \cite{Harris_2007,Jarzynski_2011,Seifert_2012}. The FT implies that while entropy production must be nonnegative on average, it may be negative in individual experimental runs. A thermodynamic description of measurement and feedback can be obtained by including an information term $\mathcal{I}$ in the FT, resulting in a generalized second law of thermodynamics that allows the average entropy production to become negative \cite{sagawa:2008}
\begin{equation}
\label{eq:secondlaw1}
    \langle \sigma\rangle \geq -\langle \mathcal{I} \rangle,
\end{equation}
where $\sigma$ denotes the entropy production. To obtain a FT, an auxiliary experimental protocol known as \textit{backward experiment} is introduced.
Interestingly, the information term is not unique. Instead, different choices of backward experiments result in FTs with different information quantifiers \cite{PottsFT2018}, which provide different bounds on the entropy production according to Eq.~\eqref{eq:secondlaw1}. 
This raises the question of which information quantifier provides the most useful constraint for feedback-controlled systems.
The two most prominent examples are: A FT where information is quantified by the mutual information or generalizations thereof \cite{SagawaGeneralized2010,horowitz:2010,ponmurugan:2010,morikuni:2011,SagawaFeedback2012,lahiri:2012,yada:2022}, and a FT that involves the entropy production that is inferable from the measurement outcomes \cite{PottsFT2018,prech:2024}. This inferable entropy is closely related to the coarse-grained entropy production that appeared in previous bounds on the entropy production \cite{kawai:2007,gomez:2008} and FTs \cite{rahav:2007,SagawaFeedback2012,kawaguchi:2013,ferri:2024}.

In addition to the FT, the entropy production enters another fundamental relation, the thermodynamic uncertainty relation (TUR) \cite{gingrich:2016,horowitz:2020}, where it bounds the signal-to-noise ratio of any current (e.g., electrical currents, particle currents, or heat currents). TURs provide insight into a variety of nano-scale systems, e.g., by constraining the efficiency of molecular motors~\cite{BaratoTUR2015} and biological clocks~\cite{MarslandTUR2019}. Through a connection between TURs and FTs \cite{timpanaro:2019,hasegawa:2019,PottsTUR2019}, generalized TURs that hold for feedback-controlled systems were recently derived \cite{PottsTUR2019,Van_VuTUR2020}. Since there are multiple information quantifiers, multiple TURs may be derived. In particular, there is a  TUR related to the mutual information and a TUR related to the inferable entropy. Just like for FTs, the question of which information quantifier results in the most useful relation is thus key for TURs as well. 
While a number of previous works have experimentally investigated FTs for feedback control \cite{toyabe_experimental_2010,koski:2014,rico:2021,debiossac:2022,kiran:2022,archambault:2024} an experimental investigation into the fluctuations of the Szilard engine energy production and its constraints has to the best of our knowledge been lacking. Indeed, experiments investigating the TUR have have only recently commenced \cite{hwang:2018,pal:2020,yang:2020,friedman:2020,manikandan:2021,shende:2025} and do not include feedback, with the exception of Refs.~\cite{paneru:2020,hazarika:2025}.

To answer these questions, we experimentally investigate the thermodynamics of information in a Szilard engine and compare different information quantifiers. To this end, we implement two different backward experiments and investigate how the thermodynamics of the Szilard engine is constrained by the corresponding information quantifiers through FTs and TURs, finding excellent agreement with theory. In particular, we verify that either FT may result in a tighter generalized second law, depending on the system parameters. For the TURs, we find that the inferable entropy is more informative than the mutual information for our experiment. It is applicable to a broader range of parameters, provides a tighter bound on the work fluctuations, and it is less affected by the finite driving speed of the Szilard engine. By explicitly implementing the backward experiments, we have access to all quantities that appear in the FT, in contrast to most previous works which only implemented the forward experiment (with Ref.~\cite{rico:2021} being a notable exception).

\begin{figure}
    \centering
    \includegraphics[width=\columnwidth]{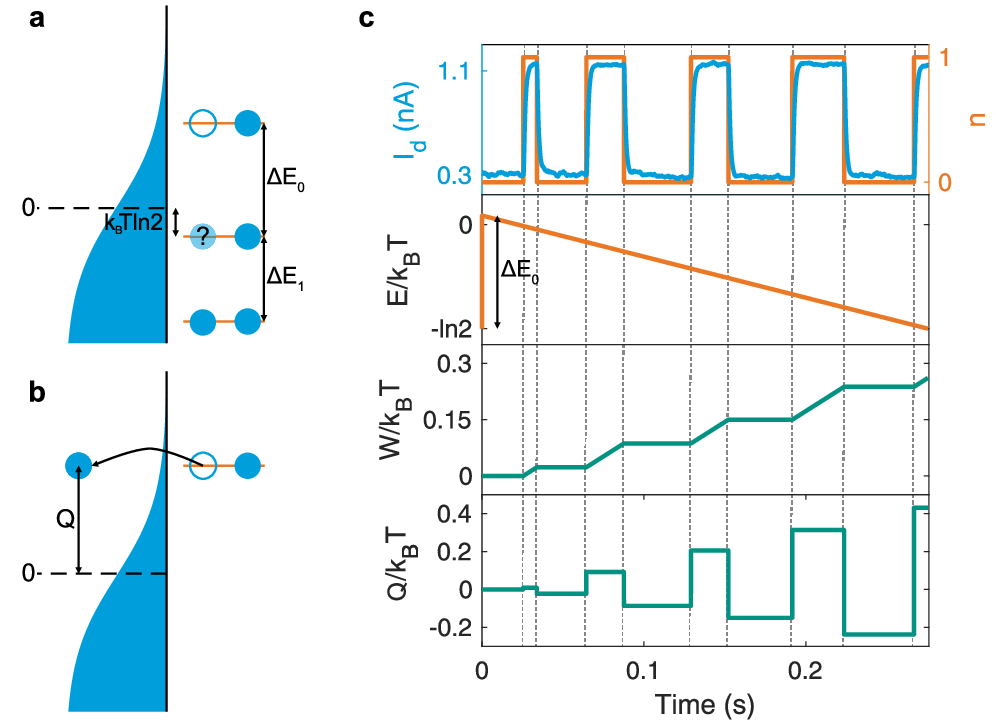}
    \caption{Szilard engine operation. a) Initially, the spin-degenerate energy level $E$ is occupied by one or two electrons with probability one half, requiring an offset from the chemical potential by $E_0 = -k_BT\ln2$. During the operation, the level is either raised by $\Delta E_0$ or lowered by $\Delta E_1$, depending on $y$, the outcome of an occupation measurement. b) The heat deposited to the reservoir when an electron tunnels into it from the quantum dot is $Q$. c) Example trajectories for $y=0$. Top panel: The charge detector current $I_d$ is used to measure the occupation $n$ of the quantum dot. Second panel: The energy level is increased by $\Delta E_0$  and then ramped back. Third panel: While $n=1$, work is extracted when $E$ is lowered. Bottom panel: Every tunneling event increases or reduces the total heat $Q$ deposited in the reservoir.}
    \label{fig:trajectory}
\end{figure}

\textit{Quantum dot Szilard engine}--Our experimental platform is a quantum dot (QD) formed in an InAs nanowire (see Ref.~\cite{BarkerFDR2022} for details on the device). The dot has a single active, spin-degenerate level at energy $E(t)$ and is tunnel coupled to an electronic reservoir at temperature $T$. The device is operated as a Szilard engine, as illustrated in Fig.~\ref{fig:trajectory}. To this end, the QD is tuned such that the active level can either contain one ($n=0$) or two electrons ($n=1$). The occupation of the energy level thus encodes one bit of information. A nearby second QD is used as an electrometer which yields a continuous real-time probe of $n(t)$, shown in Fig.~\ref{fig:trajectory}\,c). At time $t < 0$, the level is tuned such that the probability $P(n_0)=1/2$ with $n_0\equiv n(t=0)$. This corresponds to an offset from the reservoir chemical potential (set to zero for simplicity) by $E_0 = -k_BT\ln2$ due to spin degeneracy. At $t = 0$, the occupation is measured. The measurement outcome and the true occupation are denoted by $y\in\{0,1\}$ and $n_0\in\{0,1\}$ respectively. Then, a protocol that depends on the measurement outcome is applied. If $y = 0$, the level is immediately driven up by $\Delta E_0$ before being ramped back linearly to its initial value. If instead $y = 1$, $E$ is lowered by $\Delta E_1$ before being ramped back to its initial value. For both values of $y$ the time of the linear ramp is $\tau$.  
An error is made if $y\neq n_0$ which happens with an intrinsic error rate of $\epsilon \approx 0.02$. We can artificially increase $\epsilon$ by adding randomness in determining $y$. 

\begin{figure}
    \centering
    \includegraphics[width=\columnwidth]{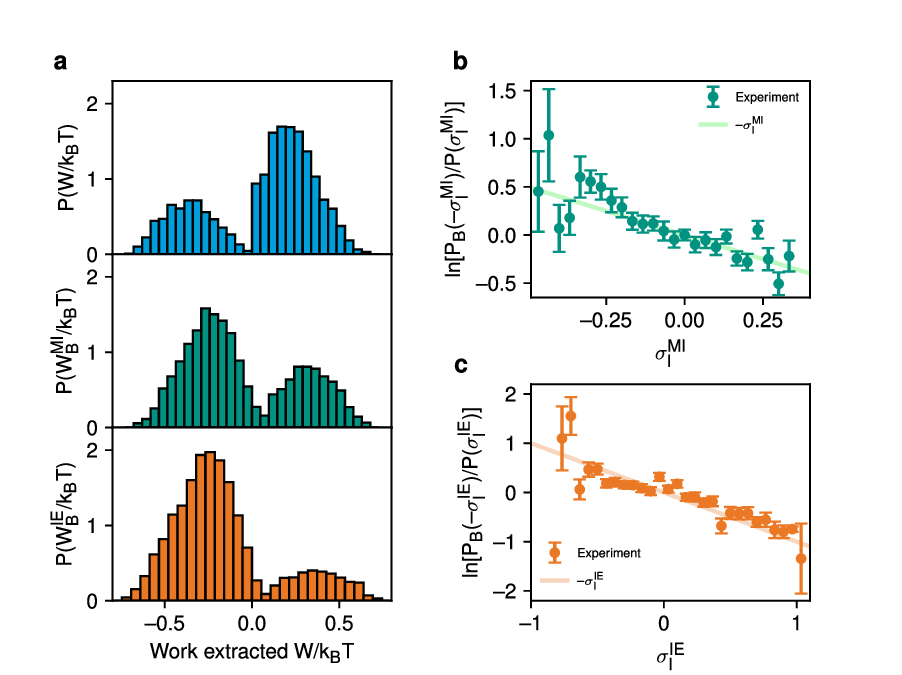}
    \caption{Fluctuation theorems. a) Example histograms of work extracted in the forward experiment (top panel) and the two different backward experiments with drive amplitudes $\Delta E_0 = \Delta E_1 = 0.75k_BT$ and error rate $\epsilon = 0.3$.
    The fluctuation theorem is verified for the backward experiment related to the mutual information in b) and the inferable entropy in c). The error bars are estimates of the standard error of the mean.}
    \label{fig:fluctuationtheorem}
\end{figure}

\textit{Stochastic thermodynamics}--The extracted work $W$ over a given trajectory $\{n(t)|0\leq t \leq \tau\}$ is given by the change in the level energy $E(t)$ while the dot is occupied, $n(t)=1$., i.e. $W \equiv -\int_0^\tau {\rm d} t \hspace{1mm} n(t) \dot{E}(t)$. Fig.~\ref{fig:trajectory}\,c) shows how $W$ is accumulated during the course of an example trajectory. The probability to obtain a certain trajectory and measurement outcome in an experimental run is given by $P(\{n(t)\},y)$. Performing many experimental runs allows us to access this distribution. In particular, we will be interested in the distribution $P(n_0,y)$, describing the initial state and the measurement outcome, as well as the work distribution $P(W)$, see Fig.~\ref{fig:fluctuationtheorem}\,a) for a histogram.

Similar to Ref.~\cite{KoskiEntropy2013}, we consider the stochastic entropy production for a given trajectory $\sigma \equiv Q/T+ \sigma_\mathrm{sys} $, where $Q $ denotes the heat that enters the reservoir whenever the dot occupation changes, illustrated in the bottom panel of Fig.~\ref{fig:trajectory}\,c). Here the entropy change in the system $\sigma_\mathrm{sys}$~\cite{SeifertEntropy2005} may be finite due to the spin-degeneracy of the $n=0$ state. For the considered scenario, we find $\sigma_\mathrm{sys} = \Delta U/T$, where $\Delta U$ is the change in internal energy. The first law of thermodynamics then implies for the total entropy production $\sigma = -W/T$, just like in the nondegenerate Szilard engine. See the supplemental material \cite{SM} for more information on the definition of the thermodynamic quantities.

\textit{Fluctation theorems}--FTs relate probabilities on forward experiments to probabilities on auxiliary experiments called backward experiments as
\begin{equation}
    \label{eq:fts}
    \frac{P_\mathrm{B}(\{n(\tau-t)\},y)}{P(\{n(t)\},y)} =e^{\frac{W(\{n(t)\},y)}{k_BT}-\mathcal{I}(n_0,y)},
\end{equation}
where $P_\mathrm{B}$ denotes the backward experiment and we anticipated that the information term $\mathcal{I}$ depends only on the measurement outcome $y$ and potentially the initial occupation $n_0$. From Eq.~\eqref{eq:fts}, the generalized second law in Eq.~\eqref{eq:secondlaw1} follows by using Jensen's inequality. In our case, it provides an upper bound on the extracted work $\langle W\rangle/k_BT \leq \langle \mathcal{I}\rangle$. Introducing the variable $\sigma_\mathrm{I}=-W/k_BT+\mathcal{I}$, Eq.~\eqref{eq:fts} can be cast into the more compact form $P_\mathrm{B}(-\sigma_\mathrm{I})/P(\sigma_\mathrm{I})=\exp(-\sigma_\mathrm{I})$ \cite{SM}.

\textit{Backward experiments}-- As pointed out in Ref.~\cite{PottsFT2018}, different choices for the backward experiment result in different FTs with different information quantifiers. These information quantifiers provide different bounds on the dissipation along the same forward experiment according to Eq.~\eqref{eq:secondlaw1}.
Here we performed two different backward experiments to investigate the thermodynamics of the Szilard engine using two information quantifiers. The forward experiment is the same in both cases and given by the operation of the Szilard engine introduced above.

The first backward experiment was introduced by Sagawa and Ueda~\cite{SagawaFeedback2012}. Instead of making a measurement, a random value for $y$ is chosen and the energy level is ramped slowly from $E_0 = -k_BT\ln2$ to the corresponding $\Delta E_y$ and then quickly returned to $E_0$. The extracted work is defined analogously to the forward protocol, see Fig.~\ref{fig:fluctuationtheorem}\,a) for a histogram of the work distribution. The information term corresponding to this backward experiment is given by
\begin{equation}
\label{eq:mist}
    \mathcal{I}^\mathrm{MI}(n_0,y) = I(n_0,y) = \ln \frac{P(n_0,y)}{P(n_0)P(y)},
\end{equation}
where in our case $P(n_0) = P(y) = 1/2$. 
Equation \eqref{eq:mist} averages to the mutual information between $n_0$ and $y$ in the forward experiment and vanishes when the measurement outcome is independent of the system state, $P(n_0,y)=P(n_0)P(y)$. In Fig.~\ref{fig:fluctuationtheorem}\,b), we show that the corresponding FT indeed holds. 

\begin{figure*}[t]
    \centering
    \includegraphics[width=\textwidth]{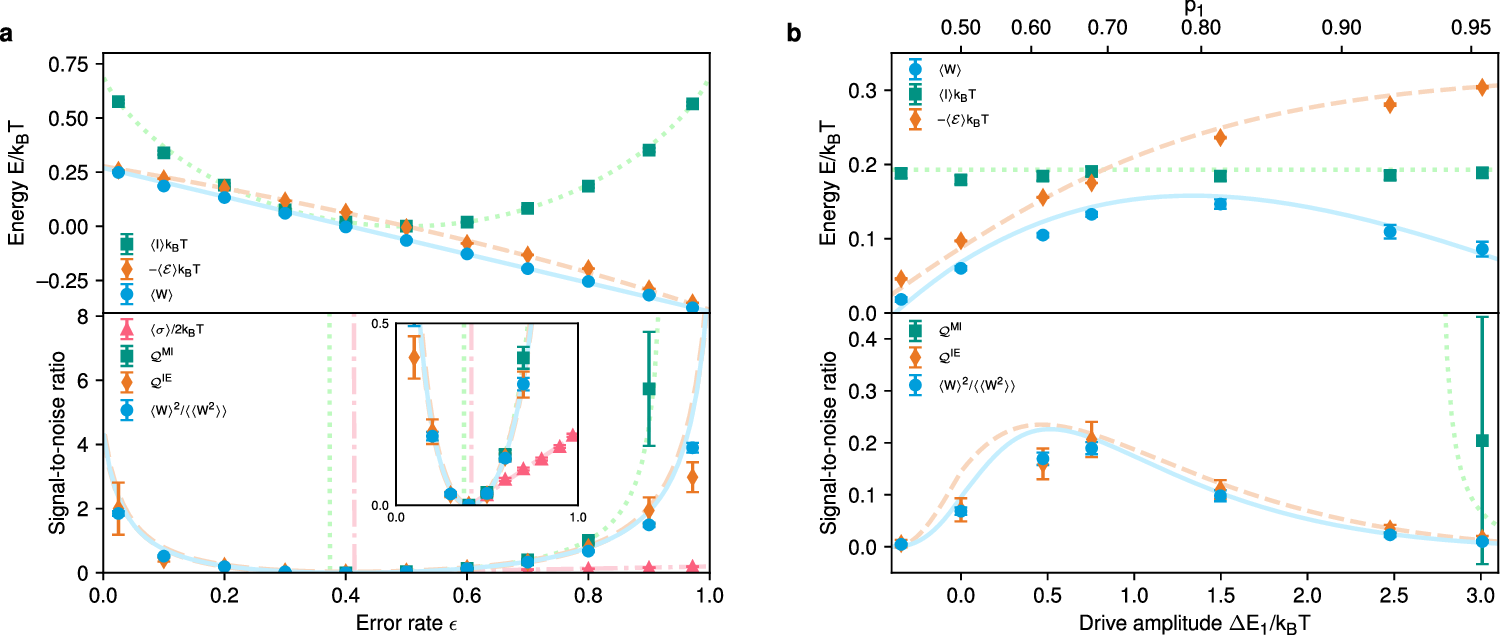}
    \caption{Generalized second law and thermodynamic uncertainty relation. Symbols are experimental data and lines correspond to theory curves. a) Top panel: The mutual information $\langle I \rangle$ (green squares) and the inferable entropy $\langle\mathcal{E}\rangle$ (orange diamonds) provide bounds on the extracted work $\langle W\rangle$ (blue circles). Bottom panel: The SNR (blue circles) are plotted against a variety of TUR quantities. Pink triangles are the RHS of the original TUR in Eq.~\eqref{eq:basicTUR}. Green squares and orange diamonds are the RHS of Eq.~\eqref{eq:GTUR}. Inset: Zooming in at the low-SNR area reveals that when $\epsilon = 0.5$, Eq.~\eqref{eq:basicTUR} is valid. Parameters: $\Delta E_0 = \Delta E_1 = 0.75k_BT$. b) Shows the same quantities as a), but now plotted against $\Delta E_1$ while $\epsilon = 0.2$ and $\Delta E_0 = 0.75k_BT$. The top horizontal axis denotes the equilibrium occupation probability at the lowest energy point. The error bars in a) and b) all represent the statistical scatter of the mean which was estimated through subsampling in the cases of the RHS of Eq.~\eqref{eq:GTUR} and $\langle\!\langle W^2\rangle\!\rangle$.}
    \label{fig:backwards}
\end{figure*}

The second backward experiment was introduced in Ref.~\cite{PottsFT2018}. 
It starts out the same as the first backward experiment but adds a final step: after the drive, the occupation is measured (with error probability $\epsilon$) and the trajectory is only kept if the outcome is equal to the randomly chosen $y$. In Fig.~\ref{fig:fluctuationtheorem}\,c) we show that this backward experiment satisfies the fluctuation theorem. In this case, the information term is (minus) the inferable entropy $\mathcal{E}$, inferable from knowing only the measurement outcome~\cite{PottsFT2018,PottsTUR2019}
\begin{equation}
\label{eq:infent}
    \mathcal{I}^\mathrm{IE}(y)=-\mathcal{E}(y) = -\ln \frac{P(y)}{P_\mathrm{S}(y)},
\end{equation}
where $P_S(y)$ is the probability that a trajectory made it through the post-selection process. 
The inferable entropy quantifies the irreversibility observed in the measurement outcome and vanishes when it is equally likely to measure $y$ in the forward and backward experiment, $P_{\rm S}(y)=P(y)$. This backward experiment is appealing since $\mathcal{E}(y)$, in contrast to $I(n_0,y)$, only depends on the measurement outcome and not on the actual occupation $n_0$. In the case of our Szilard engine, $n_0$ is accessible but that may not be true in a general measurement-feedback scenario.

The generalized second law in Eq.~\eqref{eq:secondlaw1} is verified in Fig.~\ref{fig:backwards}, where the average extracted work together with the upper bounds provided by the information terms are plotted. We find that depending on the parameters, either information quantifier may serve as the tighter bound on the extracted work. Generally, the inferable entropy provides a tight bound whenever the extracted work can be inferred from the measurement outcome. This is the case for small error rates, as well as error rates close to $\epsilon = 1$, when the outcome always reads $y\neq n_0$. We find excellent agreement between experiment and theory curves obtained by including the finite drive speed. We note that the (energy-dependent) tunnel rate, which was determined in a separate measurement, is the only free parameter in our theory \cite{SM}.

\begin{figure}
    \centering
    \includegraphics[width=\columnwidth]{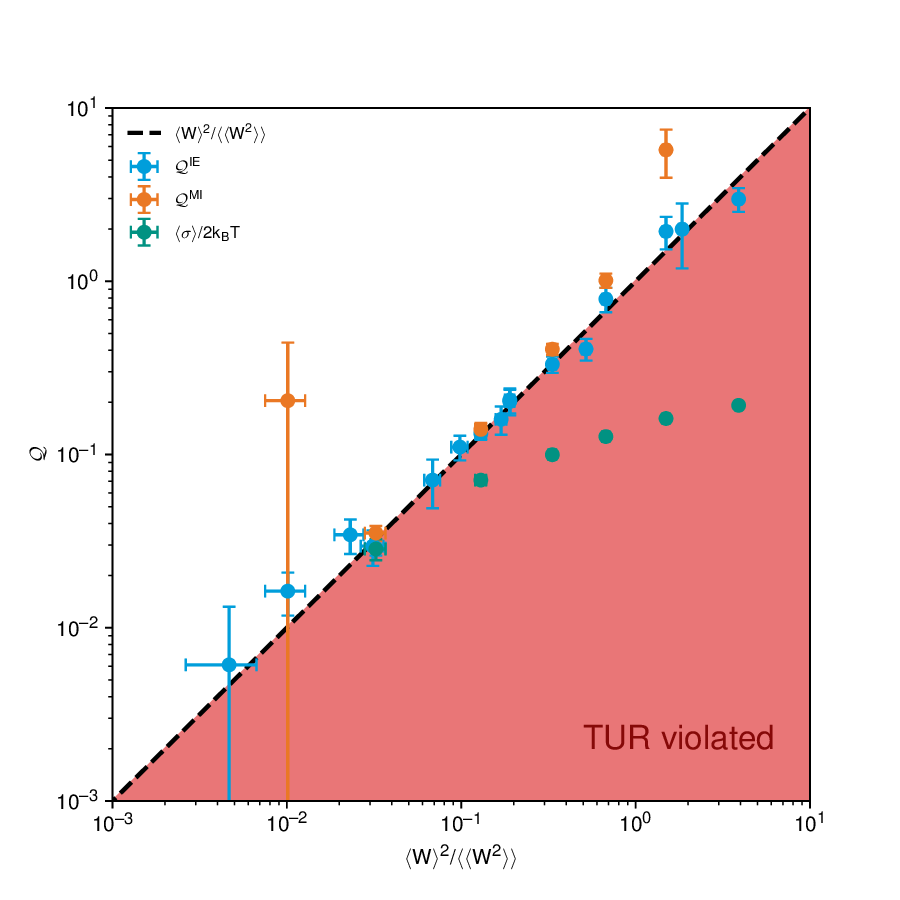}
    \caption{Thermodynamic uncertainty relations. The signal-to-noise ratio of the Szilard engine is plotted against the right-hand sides of Eqs~\eqref{eq:basicTUR} (green points) and \eqref{eq:GTUR} for two different backward experiments (blue and orange points) for the same data as in Fig.~\ref{fig:backwards}. The dashed diagonal indicates when equality is achieved. Points situated below the diagonal show the TUR in question is violated and points close to the diagonal indicate the bound is tight. All error bars show the standard error of the mean, which in the case of the blue and orange points are estimated through subsampling.}
    \label{fig:TUR}
\end{figure}

\textit{Thermodynamic uncertainty relations}--In recent years, the SNR of observables in nanoscale systems driven out of equilibrium have been shown to be bounded by the associated entropy production through TURs. The original TUR reads \cite{BaratoTUR2015}
\begin{equation}
    \label{eq:basicTUR}
    \frac{\langle W\rangle^2}{\langle\!\langle W^2\rangle\!\rangle} \leq \frac{\langle\sigma\rangle}{2k_B},
\end{equation}
where $\langle\sigma\rangle$ is the mean entropy production. 
While this relation is extremely powerful, its applicability has limitations. For instance, it is not expected to hold in cases without time-reversal symmetry making it inapplicable to general measurement and feedback scenarios. 
For such scenarios, we derive the following TUR \cite{SM}, which is a slightly tighter version of the TUR presented in Ref.~\cite{PottsTUR2019}
\begin{equation}
\label{eq:GTUR}
\begin{aligned}
    &\frac{\langle W\rangle^2}{\langle\!\langle W^2\rangle\!\rangle} \leq \mathcal{Q}^\alpha\equiv\left[\frac{1}{2}\left(1+\frac{\langle W\rangle_\mathrm{B}}{\langle W\rangle}\right)^2\times\right.\\&\left.
    \mathrm{csch}^2\left[f\left( \frac{\langle \sigma^\alpha_\mathrm{I}\rangle+\langle\sigma^\alpha_\mathrm{I}\rangle_\mathrm{B}}{4}\right)\right]-\frac{\langle\!\langle W^2\rangle\!\rangle_\mathrm{B}}{\langle W\rangle^2}\right]^{-1}.
    \end{aligned}
\end{equation}
 The inequality only holds if the right-hand side is positive. Otherwise, the TUR does not constitute a bound on the SNR. Here, $\mathrm{csch}(x) = 1/\mathrm{sinh}(x)$ is the hyperbolic cosecant and $f(x)$ is the inverse function of $x\mathrm{tanh}(x)$. The subscript B indicates that that those quantities are from the corresponding backward experiment and $\alpha =$ MI, IE, labels the backward experiment with $\sigma_{\rm I}^\alpha = -W/k_BT +\mathcal{I}^\alpha$ \cite{SM}. 
 
For the two backward experiments discussed above, Eq.~\eqref{eq:GTUR} results in two TURs that bound the SNR for the Szilard engine. Experimental results are presented in Figure~\ref{fig:backwards}, which shows the  left and right-hand sides of Eq.~\eqref{eq:GTUR} plotted against the error rate and the drive amplitude respectively. In addition, the bound provided by the original TUR, see Eq.~\eqref{eq:basicTUR}, is plotted with $\sigma=-W/T$. This bound is only applicable when $W\leq0$, i.e., when work is dissipated instead of produced by the engine. As shown in Fig.~\ref{fig:backwards}\,a), this is only the case for sufficiently high error rate. While the original TUR is obeyed for $\epsilon = 1/2$, where the engine is driven completely randomly, it is violated for larger error rates (the pink triangles are below the blue dots) because the feedback results in a much larger SNR than what is possible without feedback. Similar to the original TUR, the TUR associated to the mutual information requires a sufficiently strong error rate (or a sufficently large drive amplitude, see Fig.~\ref{fig:backwards}\,b)) to provide a bound on the SNR. For small error rates, the fluctuations on the backward experiment dominate and the right-hand side of Eq.~\eqref{eq:GTUR} becomes negative. The TUR associated to the inferable entropy is close to equality for all considered parameters (see orange and blue dots). This is further illustrated in Fig.~\ref{fig:TUR}, where $\mathcal{Q}^\alpha$ is plotted against the SNR for the different backward experiments. Including the finite driving speed in the theory was crucial for obtaining good agreement with the experimental data. This is in particular true for the TUR based on the mutual information, which is affected more strongly by the finite driving speed than the TUR based on the inferable entropy \cite{SM}. 

\textit{Conclusions}--Our results constitute a detailed investigation into the thermodynamics of information in a Szilard engine based on two different information quantifiers. Furthermore, we provide experimental tests of the TUR in a feedback scenario, where time-reversal symmetry is explicitly broken. We show how two different backward experiments provide insight into the same forward experiment through different information quantifiers. While either information quantifier may provide a tighter bound on the extracted work, we find that the inferable entropy has a number of advantages over the mutual information: It is accessible from the measurement outcomes alone and provides a tight second law whenever the extracted work is exactly known from the measurement outcomes. Furthermore, it results in a tighter TUR that is less affected by finite drive speeds.

Promising avenues for further research are provided by extending our results to other feedback scenarios in order to establish the generality of the conclusions drawn here. Furthermore, other bounds on the SNR in classical \cite{terlizzi:2019,vo:2022,prech:2025} or quantum systems \cite{carollo:2019,guarnieri:2019,hasegawa:2020,HasegawaQuantum2021,miller:2021,hasegawa:2021,vu:2022,prech:2025b} may be generalized to include feedback providing further insight into the capabilities and limitations of feedback control, see Ref.~\cite{hasegawa:2024} for a first step in this direction.  

P.P.P. acknowledges funding from the European
Union’s Horizon 2020 research and innovation programme under the Marie Sk\l odowska-Curie Grant Agreement No.~796700, from the Swedish Research Council (Starting Grant
2020-03362), and from the Swiss National Science Foundation (Eccellenza Professorial
Fellowship PCEFP2\_194268).

\bibliography{TURreferences}


\widetext
\pagebreak
\begin{center}
\textbf{\large Supplemental Material: Information Thermodynamics in an Experimentally Realized Szilard Engine: From Fluctuation Theorems to Thermodynamic Uncertainty Relations}
\end{center}
\begin{center}
David Barker,$^{1}$ Sebastian Lehmann,$^{1}$ Kimberly A. Dick,$^{1,2}$ Peter Samuelsson,$^3$ Ville F. Maisi,$^{1}$ and Patrick P. Potts$^4$
\end{center}
\begin{center}
$^1$\textit{NanoLund and Solid State Physics, Lund University, Box 118, 22100 Lund, Sweden}\\
$^2$\textit{Centre for Analysis and Synthesis, Lund University, Box 124, 22100 Lund, Sweden}\\
$^3$\textit{Physics Department and NanoLund, Lund University, Box 118, 22100 Lund, Sweden}\\
$^4$\textit{Department of Physics and Swiss Nanoscience Institute,\\ University of Basel, Klingelbergstrasse 82, 4056 Basel, Switzerland}\\
(Dated: \today)
\end{center}

\setcounter{equation}{0}
\setcounter{figure}{0}
\setcounter{table}{0}
\setcounter{page}{1}
\makeatletter
\renewcommand{\theequation}{S\arabic{equation}}
\renewcommand{\thefigure}{S\arabic{figure}}
\renewcommand{\thetable}{S\arabic{table}}

Here we provide a derivation of the thermodynamic uncertainty relation (TUR) [Eq.~\eqref{eq:GTUR}] used in the main text, we provide further details on the thermodynamic quantities in the Szilard engine, as well as on the theory used in the main text, see Fig.~\ref{fig:backwards}, and we provide details on the experiment.

\section{A. Stochastic thermodynamics of the Szilard engine}

The stochastic internal energy along a trajectory may be defined as
\begin{equation}\label{StochIntEnergy}
    U(t) \equiv n(t)E(t).
\end{equation}
To determine heat and work, we examine the change in the internal energy in a time-step of ininitesimal duration $dt$. To this end, we first consider the change in $dn(t) = n(t+dt)-n(t)$. Either, the occupation stays the same, $dn=0$, or an electron is exchanged with the reservoir $dn=\pm 1$. The incremental change in the internal energy is thus
\begin{equation}
    dU(t) = \dot{E}(t) n(t) dt + E(t) dn(t),
\end{equation}
where we identify the first term with power, while the second term represents a heat current. We thus arrive at the definitions for stochastic heat and work
\begin{equation}
\label{eq:work}
    W = -\int_0^\tau dt \dot{E}(t) n(t), \hspace{2cm} Q = -\int_{\{n(t)\}}dn(t)\, E(t),
\end{equation}
where the integral in the heat is a stochastic integral along the trajectory $\{n(t)|0\leq t\leq \tau\}$. We note that work is positive if it is extracted and heat is positive if it enters the reservoir. The first law of thermodynamics then reads
\begin{equation}\label{1stLaw}
    \Delta{U}\equiv U(\tau)-U(0) = -{W} - {Q}.
\end{equation}

The entropy production along a trajectory is given by
\begin{equation}
    \label{eq:stochent}
    \sigma = Q/T + \sigma_\mathrm{sys}.
\end{equation}
To determine the entropy change in the system, $\sigma_\mathrm{sys}$, we note that our Szilard engine is based on a three level system. The $n=0$ state, where a single electron with either spin up or down is on the dot, is doubly degenerate state and has zero energy by convention. The $n=1$ state, where two electrons are on the dot, is non-degenerate and has energy $E_0$ (at time $t\leq 0$). The system thus starts in the equilibrium state characterized by the probabilities
\begin{equation}
    \label{eq:eq}
    p_{0,\uparrow} = p_{0,\downarrow} = \frac{1}{2+e^{-E_0/(k_BT)}},\hspace{1cm}p_1 = \frac{e^{-E_0/(k_BT)}}{2+e^{-E_0/(k_BT)}}.
    \end{equation}
    For $E_0 = -k_BT\ln 2$, we find $p_1=1/2$, and $p_{0,\uparrow} = p_{0,\downarrow}=1/4$. The system entropy may thus take on three different values
    \begin{equation}
        \label{eq:sysent}
        \sigma_\mathrm{sys} = k_B\ln \frac{p_{n_0}}{p_{n(\tau)}} = \begin{cases}
            0 \hspace{.5cm}\text{for}\hspace{.5cm} n_0=n(\tau),\\
            k_B\ln 2 \hspace{.5cm}\text{for}\hspace{.5cm} n_0=1,\,n(\tau)=0,\\
            -k_B \ln 2  \hspace{.5cm}\text{for}\hspace{.5cm} n_0=0,\,n(\tau)=1,
        \end{cases}
    \end{equation}
    where we abbreviated $p_0 = p_{0,\uparrow} = p_{0,\downarrow}$.
Similarly, the internal energy change may take on three different values
  \begin{equation}
        \label{eq:Usys}
        \Delta U =  \begin{cases}
            0 \hspace{.5cm}\text{for}\hspace{.5cm} n_0=n(\tau),\\
            -E_0 =  k_BT \ln 2\hspace{.5cm}\text{for}\hspace{.5cm} n_0=1,\,n(\tau)=0,\\
            E_0 = -k_BT \ln 2  \hspace{.5cm}\text{for}\hspace{.5cm} n_0=0,\,n(\tau)=1.
        \end{cases}
    \end{equation}
We thus conclude that $\sigma_\mathrm{sys}=\Delta U/T$. Inserting this into Eq.~\eqref{eq:stochent}, and using the first law in Eq.~\eqref{1stLaw}, we find $\sigma = -W/T$.

\section{B. Derivation of the TUR}
To derive Eq.~\eqref{eq:GTUR} in the main text, we write the fluctuation theorem (FT) in Eq.~\eqref{eq:fts} as
\begin{equation}
\label{eq:flucrel}
\frac{P_{\rm B}(-W,-\sigma_{\rm I})}{P(W,\sigma_{\rm I})}=e^{-\sigma_{\rm I}}.
\end{equation}
To this end, we introduced
\begin{equation}
\label{eq:forwack}
    \begin{aligned}
    &P(W,\sigma_\mathrm{I}) = \sum_{\{n(t)\},y} \delta\big(W-W(\{n(t),y\})\big)\delta\big(\sigma_\mathrm{I}+W/(k_BT)-\mathcal{I}(n_0,y)\big)P(\{n(t)\},y),\\
    &P_\mathrm{B}(W,\sigma_\mathrm{I}) = \sum_{\{n(t)\},y} \delta\big(W-W(\{n(t),y\})\big)\delta\big(\sigma_\mathrm{I}+W/(k_BT)+\mathcal{I}(n(\tau),y)\big)P_\mathrm{B}(\{n(t)\},y).
    \end{aligned}
\end{equation}
Equation \eqref{eq:flucrel} may then be derived by re-writing Eq.~\eqref{eq:fts} in the main text as
\begin{equation}
    \label{eq:ftder1}
     P_\mathrm{B}(\{n(\tau-t)\},y) =e^{\frac{W(\{n(t)\},y)}{k_BT}-\mathcal{I}(n_0,y)}P(\{n(t)\},y).
\end{equation}
We now multiply this equation by $\delta\big(W-W(\{n(t),y\})\big)\delta\big(\sigma_\mathrm{I}+W/(k_BT)-\mathcal{I}(n_0,y)\big)$ and sum over all trajectories $\{n(t)\}$ and all measurement outcomes. The right-hand side then results in
 \begin{equation}
 \label{eq:ftderfor}
 \begin{aligned}
 &\sum_{\{n(t)\},y} \delta\big(W-W(\{n(t),y\})\big)\delta\big(\sigma_\mathrm{I}+W/(k_BT)-\mathcal{I}(n_0,y)\big) e^{\frac{W(\{n(t)\},y)}{k_BT}-\mathcal{I}(n_0,y)}P(\{n(t)\},y)\\&=e^{-\sigma_\mathrm{I}}\sum_{\{n(t)\},y} \delta\big(W-W(\{n(t),y\})\big)\delta\big(\sigma_\mathrm{I}+W/(k_BT)-\mathcal{I}(n_0,y)\big) P(\{n(t)\},y)=e^{-\sigma_\mathrm{I}}P(W,\sigma_\mathrm{I}).
 \end{aligned}
 \end{equation}
 The left-hand side of Eq.~\eqref{eq:ftder1} results in
 \begin{equation}
 \label{eq:ftderback}
 \begin{aligned}
    & \sum_{\{n(t)\},y} \delta\big(W-W(\{n(t),y\})\big)\delta\big(\sigma_\mathrm{I}+W/(k_BT)-\mathcal{I}(n_0,y)\big)P_\mathrm{B}(\{n(\tau-t)\},y) \\&= \sum_{\{n(t)\},y} \delta\big(W-W(\{n(\tau-t),y\})\big)\delta\big(\sigma_\mathrm{I}+W/(k_BT)-\mathcal{I}(n(\tau),y)\big)P_\mathrm{B}(\{n(t)\},y) \\& = \sum_{\{n(t)\},y} \delta\big(W+W(\{n(t),y\})\big)\delta\big(\sigma_\mathrm{I}+W/(k_BT)-\mathcal{I}(n(\tau),y)\big)P_\mathrm{B}(\{n(t)\},y) = P(-W,-\sigma_\mathrm{I}),
     \end{aligned}
 \end{equation}
 where in the first equality, we used the fact that we may instead over $\{n(t)\}$ also sum over the time-reversed trajectories $\{n(\tau-t)\}$ and therefore we may time-reverse all trajectories under the sum. In the second equality, we further used that work is odd under time-reversal, i.e., $W(\{n(\tau-t)\})=-W(\{n(t)\})$. Setting Eq.~\eqref{eq:ftderfor} equal to Eq.~\eqref{eq:ftderback}, and dividing by $P(W,\sigma_\mathrm{I})$, we arrive at Eq.~\eqref{eq:flucrel}. Note that Eq.~\eqref{eq:forwack} implies that the work associated to the backward experiment is indeed the physical work as defined above in Eq.~\eqref{eq:work}. The information assigned to a backward experiment is minus the information that we associated to the corresponding forward experiment. This choice ensures that $\sigma_\mathrm{I}$ is odd under time-reversal. As discussed below, we do not need to assign an information to the backward experiment to evaluate the TUR. We note that integrating Eqs.~\eqref{eq:ftderfor} and \eqref{eq:ftderback} over $W$, we can derive the FT
 \begin{equation}
     \label{eq:ftplot}
     \frac{P_{\rm B}(-\sigma_{\rm I})}{P(\sigma_{\rm I})}=e^{-\sigma_{\rm I}},
 \end{equation}
 which is plotted in Fig.~\ref{fig:fluctuationtheorem} in the main text.

Following the appendix of Ref.~\cite{Van_VuTUR2020}, we can obtain a TUR including measurement and feedback that is tighter than the one presented in Ref.~\cite{PottsTUR2019}. To this end, we introduce the auxiliary probability distribution
\begin{equation}
\label{eq:qprob}
Q(W,\sigma_{\rm I}) =\frac{1}{2}\left[P(W,\sigma_{\rm I})+P_{\rm B}(-W,-\sigma_{\rm I})\right]= \frac{1+e^{-\sigma_{\rm I}}}{2}P(W,\sigma_{\rm I}),
\end{equation}
which is the average of the forward distribution and the backward distribution with negated arguments, and we used Eq.~\eqref{eq:flucrel} in the last equality. Note that $Q$ is not restricted to positive $\sigma_{\rm I}$. An important property of $Q$ that will be used repeatedly is given by
\begin{equation}
\tanh\left(\frac{\sigma_{\rm I}}{2}\right)Q(W,\sigma_{\rm I})=\frac{1}{2}\left[P(W,\sigma_{\rm I})-P_{\rm B}(-W,-\sigma_{\rm I})\right].
\end{equation}
We now prove the series of inequalities
\begin{equation}
\label{eq:ineqsder}
\begin{aligned}
\left(\frac{\langle W\rangle+\langle W\rangle_{\rm B}}{2}\right)^2&=\left\langle\left(W-\langle W\rangle_Q\right)\tanh\left(\frac{\sigma_{\rm I}}{2}\right)\right\rangle_Q^2
\\&\leq \langle\!\langle W^2\rangle\!\rangle_Q\left\langle \tanh^2\left(\frac{\sigma_{\rm I}}{2}\right)\right\rangle_Q=\langle\!\langle W^2\rangle\!\rangle_Q\left\langle \tanh^2\left[f\left(\frac{\sigma_{\rm I}}{2}\tanh\left(\frac{\sigma_{\rm I}}{2}\right)\right)\right]\right\rangle_Q\\&\leq\langle\!\langle W^2\rangle\!\rangle_Q \tanh^2\left[f\left(\left\langle\frac{\sigma_{\rm I}}{2}\tanh\left(\frac{\sigma_{\rm I}}{2}\right)\right\rangle_Q\right)\right]=\langle\!\langle W^2\rangle\!\rangle_Q \tanh^2\left[f\left(\frac{\langle\sigma_{\rm I}\rangle+\langle\sigma_{\rm I}\rangle_{\rm B}}{4}\right)\right].
\end{aligned}
\end{equation}
where we introduced the average over $Q$ as $\langle\cdot\rangle_Q$, the average over $P$ as $\langle\cdot\rangle$ and the average over $P_{\rm B}$ as $\langle\cdot\rangle_{\rm B}$.
The first inequality is the Cauchy-Schwarz inequality. In the following equality, we introduced $f$ as the inverse of $x\tanh(x)$, i.e., $f[x\tanh(x)]=x$. The second inequality is Jensen's inequality which holds because $\tanh^2[f(x)]$ is a concave function for $x\geq0$. The following equality uses
\begin{equation}
\langle\sigma_{\rm I}\rangle+\langle\sigma_{\rm I}\rangle_{\rm B}=2\langle \sigma_{\rm I}\tanh(\sigma_{\rm I}/2)\rangle_Q.
\end{equation}
By further using
\begin{equation}
\label{eq:varphiq}
\langle\!\langle W^2\rangle\!\rangle_Q =\frac{1}{2}\left(\langle\!\langle W^2\rangle\!\rangle+\langle\!\langle W^2\rangle\!\rangle_{\rm B}\right)+\left(\frac{\langle W\rangle+\langle W\rangle_{\rm B}}{2}\right)^2,
\end{equation}
we finally obtain
\begin{equation}
\label{eq:gturfb}
\frac{\langle\!\langle W^2\rangle\!\rangle+\langle\!\langle W^2\rangle\!\rangle_{\rm B}}{\left(\langle W\rangle+\langle W\rangle_{\rm B}\right)^2}\geq\frac{1}{2}{\rm csch}^2\left[f\left(\frac{\langle\sigma_{\rm I}\rangle+\langle\sigma_{\rm I}\rangle_{\rm B}}{4}\right)\right]=\frac{1}{2}{\rm csch}^2\left[f\left(\frac{\langle\sigma_{\rm I}(1-e^{-\sigma_\mathrm{I}})\rangle}{4}\right)\right],
\end{equation}
with the hyperbolic cosecant ${\rm csch}(x)=1/\sinh(x)$. In the last equality, we wrote the right-hand side using only an average over the forward experiment, to illustrate that only the statistics of work are needed in the backward experiment, which have a clear physical interpretation. Solving the last inequality for the signal-to-noise ratio, we obtain Eq.~\eqref{eq:GTUR} in the main text.

We note that for identical forward and backward distributions, this bound reduces to the tightest bound that can be inferred from the FT \cite{timpanaro:2019}.

\section{C. Theory}
In this section, we present the theory that was used in Fig.~\ref{fig:backwards} in the main text.

\subsection{C. 1. Quasistatic Szilard engine}
We start by considering the work in the forward experiment, which depends on two quantities. The initial state $n_0$, and the measurement otucome $y$. From Eq.~\eqref{eq:work}, we may write the work in the quasi-static approximation as
\begin{equation}
    \label{eq:workconc}
    W(n_0,y) = \delta_{n_0,1}(-1)^{\delta_{y,0}}\Delta E_y+\int_0^\tau dt \dot{E}_y(t) p_1(E_y(t)) dt = \delta_{n_0,1}(-1)^{\delta_{y,0}}\Delta E_y+\int_{E_y(0)}^{E_0} p_1(E) dE, 
\end{equation}
with
\begin{equation}
    E_y(t) = \frac{t}{\tau}E_0 +  \left(1-\frac{t}{\tau}\right)(-1)^{\delta_{y,1}}\Delta E_y.
\end{equation}
In Eq.~\eqref{eq:workconc}, we used the quasi-static approximation to replace $n(t)$ with the steady-state occupation [see Ref.~\cite{PottsTUR2019} for a more detailed justification]
\begin{equation}
    p_1(E) = \frac{e^{-E/(k_BT)}}{2+e^{-E/(k_BT)}} =f(E-E_0),
\end{equation}
where $f(E) = 1/(e^{E/k_BT}+1)$ denotes the Fermi-Dirac occupation. Using $f(-E)=1-f(E)$, we may re-write Eq.~\eqref{eq:workconc} as
\begin{equation}
    \label{eq:workconc2}
    W(n_0,y) = \int_0^{\Delta E_y}f(E) dE - (1-\delta_{n_0,y})\Delta E_y =k_BT\ln 2 +\delta_{n_0,y}k_BT\ln\left[1-f(\Delta E_y)\right]+(1-\delta_{n_0,y})k_BT\ln\left[f(\Delta E_y)\right],
\end{equation}
where we used
\begin{equation}
    \int_0^\varepsilon f(E)dE = k_BT\ln 2+k_B T \ln\left[1-f(\varepsilon)\right],\hspace{2cm}k_B T \ln\left[1-f(\varepsilon)\right]=\varepsilon+k_B T \ln\left[f(\varepsilon)\right].
\end{equation}
The work in Eq.~\eqref{eq:workconc2} has exactly the same form as in the standard Szilard engine, see Eq.~(B3) in Ref.~\cite{PottsTUR2019}, but with the quantity $1-f(\Delta E_y)$ playing the role of the final fraction of the volume the particle has at its disposal, $v_y$. In the ideal Szilard engine, the the particle tends to occupy the full volume at the end ($v_y=1$), which corresponds in our case to $\Delta E_y \to \infty$, where the work of $k_BT \ln 2$ can be extracted.

The probability distribution for $n_0$ and $y$ in the forward experiment reads
\begin{equation}
    \label{eq:probfor}
    P(n_0,y) = \frac{1-\epsilon}{2}\delta_{n_0,y}+\frac{\epsilon}{2}(1-\delta_{n_0,y}).
\end{equation}
This results in the average work
\begin{equation}
    \label{eq:avwork}
    \beta\langle W\rangle = \ln 2+\sum_{y=0,1} \left\{\frac{1-\epsilon}{2}\ln\left[1-f(\Delta E_y)\right]+\frac{\epsilon}{2}\ln\left[f(\Delta E_y)\right]\right\},
\end{equation}
and its second moment
\begin{equation}
    \label{eq:varwork}
    \beta^2\langle W^2\rangle = \sum_{y=0,1} \left\{\frac{1-\epsilon}{2}\ln^2\left[2-2f(\Delta E_y)\right]+\frac{\epsilon}{2}\ln\left[2f(\Delta E_y)\right]\right\},
\end{equation}
with $\langle\!\langle W^2\rangle\!\rangle=\langle W^2\rangle-\langle W\rangle^2$ and $\beta = 1/k_BT$.

The information term in Eq.~\eqref{eq:mist} reads
\begin{equation}
    I(n_0,y) = \ln 2 +\delta_{n_0,y}\ln(1-\epsilon)+(1-\delta_{n_0,y})\ln \epsilon,
\end{equation}
which averages to the mutual information
\begin{equation}
    \label{eq:mutint}
    \langle I \rangle = \ln 2 +\frac{1}{2}\sum_{y=0,1}\left[\epsilon \ln \epsilon+(1-\epsilon)\ln (1-\epsilon)\right].
\end{equation}
The backward experiment associated to the mutual information is described by the probability distribution
\begin{equation}
    \label{eq:backwardmi}
    P_{\mathrm{B}}^{\mathrm{MI}}(n_\tau,y) = \frac{1}{2}\delta_{n_\tau,y}[1-f(\Delta E_y)]+\frac{1}{2}(1-\delta_{n_\tau,y})f(\Delta E_y).
\end{equation}
As discussed in the main text, no measurement is performed, instead a value for $y$ is chosen with probability $1/2$. The final occupation of the dot, $n_\tau\equiv n(\tau)$, is then, due to the quasi-static assumption, determined by the Fermi-Dirac occupation at the final energy of the linear drive. From the distribution in Eq.~\eqref{eq:backwardmi}, we may derive
\begin{equation}
    \beta\langle W\rangle_\mathrm{B} = -\beta\sum_{n_\tau,y}P_{\mathrm{B}}^{\mathrm{MI}}(n_\tau,y) W(n_\tau,y)= -\ln 2 - \frac{1}{2}\sum_{y=0,1} \left\{\left[1-f(\Delta E_y)\right]\ln\left[1-f(\Delta E_y)\right]+f(\Delta E_y)\ln\left[f(\Delta E_y)\right]\right\},
\end{equation}
\begin{equation}
    \beta^2\langle W^2\rangle_\mathrm{B} = \beta^2\sum_{n_\tau,y}P_{\mathrm{B}}^{\mathrm{MI}}(n_\tau,y) W^2(n_\tau,y)= \frac{1}{2}\sum_{y=0,1} \left\{\left[1-f(\Delta E_y)\right]\ln^2\left[2-2f(\Delta E_y)\right]+f(\Delta E_y)\ln^2\left[2f(\Delta E_y)\right]\right\},
\end{equation}
and
\begin{equation}
    \langle I\rangle_\mathrm{B} = -\sum_{n_\tau,y}P_{\mathrm{B}}^{\mathrm{MI}}(n_\tau,y) I(n_\tau,y) = -\ln 2 -\frac{1}{2}\sum_{y=0,1} \left\{\left[1-f(\Delta E_y)\right]\ln(1-\epsilon)+f(\Delta E_y)\ln \epsilon\right\}.
\end{equation}

The second backward experiment adds a post-selection step to the first one. To this end, a measurement is performed with error probability $\epsilon$. This multiplies the term in Eq.~\eqref{eq:backwardmi} where $n_\tau=y$ with the probability of not making a mistake $(1-\epsilon)$. Similarly, the term where $n_\tau\neq y$ is multiplied by $\epsilon$. Finally, the postselection ensures that the distribution is again normalized resulting in
\begin{equation}
    \label{eq:backwardie}
    P_{\mathrm{B}}^{\mathrm{IE}}(n_\tau,y) = \frac{1}{2}\frac{\delta_{n_\tau,y}[1-f(\Delta E_y)](1-\epsilon)+(1-\delta_{n_\tau,y})f(\Delta E_y)\epsilon}{[1-f(\Delta E_y)](1-\epsilon)+f(\Delta E_y)\epsilon}.
\end{equation}
The success probability in Eq.~\eqref{eq:infent} in the main text is given by the normalization factor
\begin{equation}
    P_\mathrm{S}(y) = [1-f(\Delta E_y)](1-\epsilon)+f(\Delta E_y)\epsilon,
\end{equation}
resulting in the inferable entropy
\begin{equation}
    \mathcal{E}(y) = -\ln 2-\ln\left\{[1-f(\Delta E_y)](1-\epsilon)+f(\Delta E_y)\epsilon\right\}.
\end{equation}
For the second backward experiment, we find
\begin{equation}
    \beta\langle W\rangle_\mathrm{B} = -\beta\sum_{n_\tau,y}P_{\mathrm{B}}^{\mathrm{IE}}(n_\tau,y) W(n_\tau,y)= -\ln 2 - \frac{1}{2}\sum_{y=0,1} \frac{(1-\epsilon)\left[1-f(\Delta E_y)\right]\ln\left[1-f(\Delta E_y)\right]+\epsilon f(\Delta E_y)\ln\left[f(\Delta E_y)\right]}{[1-f(\Delta E_y)](1-\epsilon)+f(\Delta E_y)\epsilon},
\end{equation}
\begin{equation}
    \beta^2\langle W^2\rangle_\mathrm{B} = \beta^2\sum_{n_\tau,y}P_{\mathrm{B}}^{\mathrm{IE}}(n_\tau,y) W^2(n_\tau,y)= \frac{1}{2}\sum_{y=0,1} \frac{(1-\epsilon)\left[1-f(\Delta E_y)\right]\ln^2\left[2-2f(\Delta E_y)\right]+\epsilon f(\Delta E_y)\ln^2\left[2f(\Delta E_y)\right]}{[1-f(\Delta E_y)](1-\epsilon)+f(\Delta E_y)\epsilon},
\end{equation}
and
\begin{equation}
   \langle \mathcal{E}\rangle= -\langle \mathcal{E}\rangle_\mathrm{B} = \sum_{n_\tau,y}P_{\mathrm{B}}^{\mathrm{IE}}(n_\tau,y) \mathcal{E}(y) = -\ln 2 -\frac{1}{2}\sum_{y=0,1} \ln\left\{\left[1-f(\Delta E_y)\right](1-\epsilon)+f(\Delta E_y) \epsilon\right\}.
\end{equation}

\subsection{C. 2. Finite driving speed}
To capture corrections due to a finite driving speed, we follow Ref.~\cite{BarkerFDR2022}.
The dynamics of the quantum dot is governed by the equation
\begin{equation}
    \label{eq:master}
    \partial_t q_1 = \Gamma f(E)(1-q_1)-2\Gamma[1-f(E)]q_1, 
\end{equation}
where $q_1$ is the probability of the dot being occupied ($n=1$), $\Gamma$ is the tunnel rate that may depend on $E$ (and thus on time), see Sec.~D. 3. below. The factor of $2$ in front of the tunneling-out rate is due to the degeneracy of the $n=0$ state. Introducing the dimensionless time $s=t/\tau$, we may write
\begin{equation}
    \label{eq:master2}
    \frac{1}{\tau}\partial_sq_1 = \Gamma[2-f(E)]\left[p_1(E)-q_1\right].
\end{equation}
We now expand 
\begin{equation}
    \label{eq:expansion}
    q_1 = q_1^{(0)}+\frac{1}{\tau}q_1^{(1)}+\cdots
\end{equation}
and solve Eq.~\eqref{eq:master2} order by order, finding
\begin{equation}
    q_1^{(0)} = p_1(E),\hspace{2cm}q_1^{(1)} = -\frac{1}{\Gamma[2-f(E)]}\partial_s p_1(E),
\end{equation}
which results in
\begin{equation}
    \label{eq:probapprox}
    q_1(t) \simeq p_1[E(t)]-\frac{1}{\tau\Gamma[2-f(E)]}\partial_t p_1[E(t)] = p_1[E(t)]+\frac{\beta \dot{E}(t)}{\tau\Gamma[2-f(E)]}p_1[E(t)]\left\{1-p_1[E(t)]\right\}.
\end{equation}

From Eq.~\eqref{eq:work} we find $\langle W\rangle = \langle W\rangle_\mathrm{inf}-W_\mathrm{diss} $, where $\langle W\rangle_\mathrm{inf}$ denotes the work extracted at infinitely slow drives, as discussed in the last subsection, and the dissipated work is given by
\begin{equation}
    \label{eq:wdiss}
    W_\mathrm{diss} =  \frac{\beta}{\tau}\int_0^\tau dt \frac{[\dot{E}(t)]^2}{\Gamma[2-f(E)]}p_1[E(t)]\left\{1-p_1[E(t)]\right\}.
    \end{equation}
    The correction to the variance in the extracted work is determined by the work fluctuation-dissipation relation~\cite{hermans:1991,jarzynski:1997,BarkerFDR2022}
    \begin{equation}
        \langle\!\langle W^2\rangle\!\rangle = \langle\!\langle W^2\rangle\!\rangle_{\mathrm{inf}}+2k_BTW_{\mathrm{diss}}.
    \end{equation}

\begin{figure}[h]
    \centering
    \includegraphics[width = \linewidth]{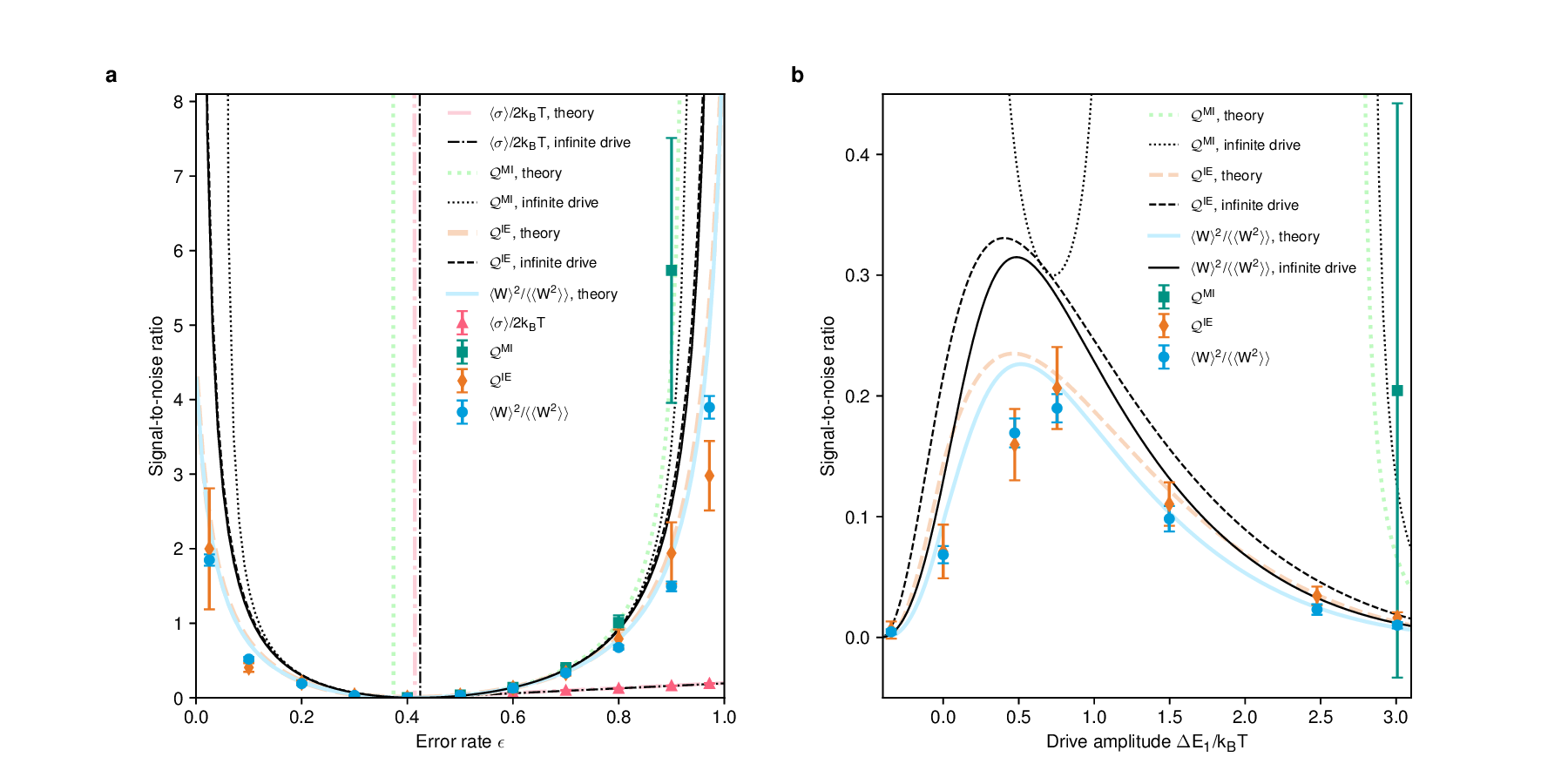}
    \caption{The same experimental data and full theory curves from the bottom panels of Figure~\ref{fig:backwards} (a) and (b) in the main text. The additional black lines are theory curves that neglect the effect of the finite driving speed of the protocol.}
    \label{fig:infinite_drive}
\end{figure}

    Figure \ref{fig:infinite_drive} illustrates the contribution of the finite drive speed on the TUR. For the TUR associated to the mutual information, the finite drive speed renders the RHS of Eq.~\eqref{eq:GTUR} negative for a range of parameters (e.g., small error rates), reducing the applicability of the corresponding TUR.

\section{D. Experimental details}
The device and measurement setup are the same as in Ref.~\cite{BarkerFDR2022}, to which we refer for a detailed description. Below we briefly summarize the fabrication and measurements for completeness as well as describe the measurement of the energy-dependent tunneling rate used to theoretically calculate the dissipated work in Eq.~(\ref{eq:wdiss}).
\subsection{D. 1. Device fabrication}

\begin{figure}[h]
    \centering
    \includegraphics[width=0.6\linewidth]{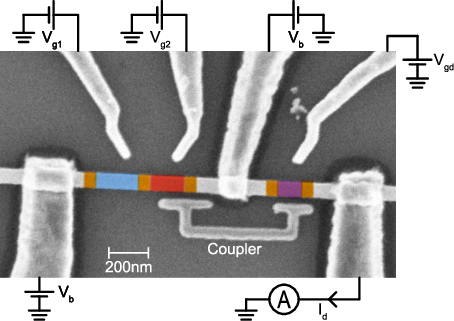}
    \caption{False-color SEM micrograph of the device used in the study. Orange sections indicate the tunnel barriers which form three quantum dots. The blue section contains the QD under study, the red section contains a QD which is put into Coulomb blockade, and the purple section is the QD used for charge sensing. Figure reproduced from Ref.~\cite{BarkerFDR2022}.}
    \label{fig:device}
\end{figure}

The device, shown in Figure~\ref{fig:device}, was fabricated from an InAs nanowire grown by metal-organic vapor phase epitaxy, with three quantum dots formed by polytype engineering as described in Refs.~\cite{LehmannCrystalPhase2013, NamaziGaSb2015,BarkerDQD2019}. Electrical leads were fabricated using electron beam lithography. A central lead divides the three-dot system into two sections: a double quantum dot (DQD), and a single quantum dot (SQD). In addition to the central lead, each section has a dedicated lead. Each quantum dot is controlled by a nearby gate electrode fabricated in a separate lithography step. A metal strip, deposited along with the gates, provides capacitive coupling between the two sections. 
\subsection{D. 2. Measurements} 
All measurements in this study were performed in a dilution refrigerator at \SI{100}{\milli\kelvin}. A bias voltage of $V_{b} =$~\SI{1}{\milli\volt} was applied to the SQD section, which was tuned with the gate voltage $V_{gd}$ such that the resulting current $I_d$ was sensitive to changes in the charge state of the DQD. As in Ref.~\cite{BarkerFDR2022}, an Arduino DUE board provided the measurement and feedback control, and the signal was captured using a data-acquisition card. For more details on how the signal was amplified, filtered, and processed to produce a time trace of the occupancy $n(t)$, we again refer to the Supplemental Materials of that reference.
\\ \\In each forward experiment, $5000$ trajectories were collected, with the corresponding measurement outcome $y$ recorded for each run. For the backward experiment based on mutual information, trajectories were sampled to match the distribution of $y$ observed in the forward process. In the case involving inferrable entropy, post-selection was applied: a larger initial set of trajectories was acquired, and unmatched outcomes were discarded to yield $5000$ trajectories with the same $y$-distribution as the forward case. 
\subsection{D. 3. Tunnel rates and lever arm}
\begin{figure}[h]
    \centering
    \includegraphics[width = \linewidth]{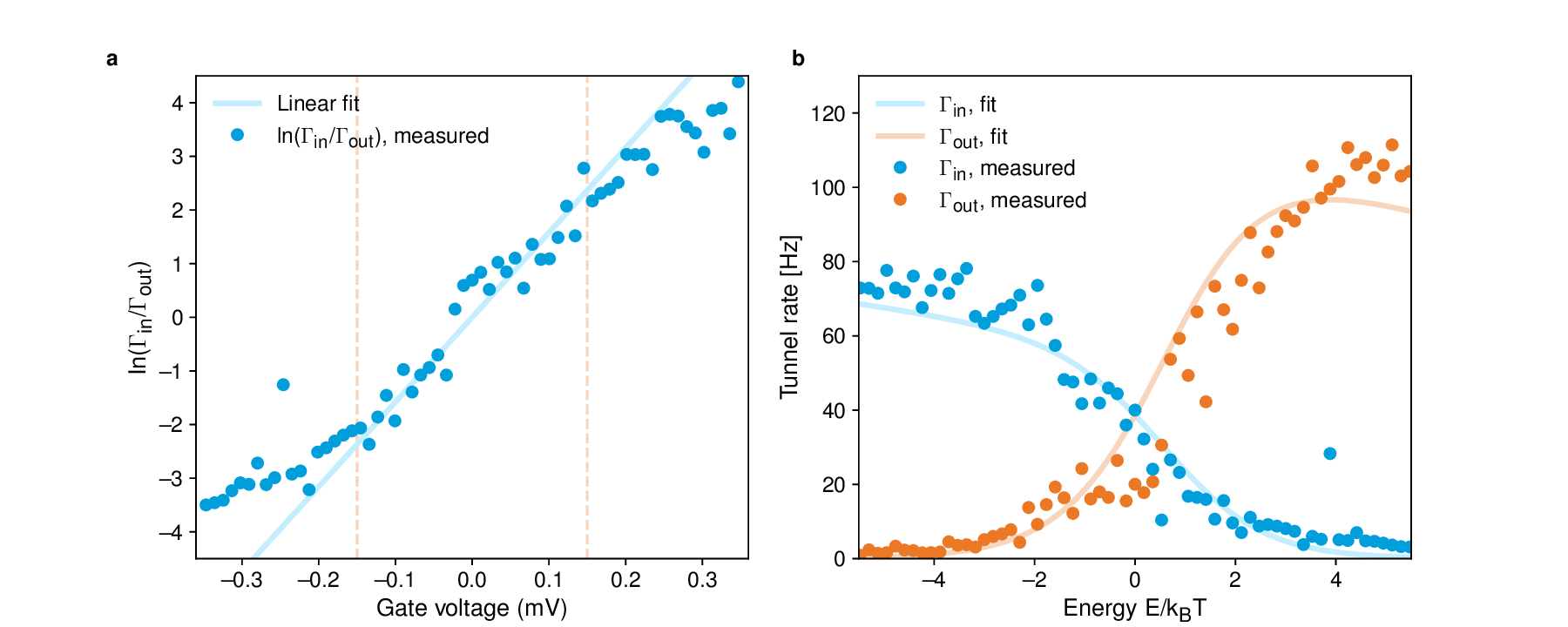}
    \caption{a) Plotting the the gate voltage $V_{g1}$ against $\mathrm{ln}(\Gamma_{in}/\Gamma_{out})$ illustrates that the tunnel rates follow a detailed balanced condition. The light blue line is a linear fit to experimental data, from which the lever arm $\alpha$ can be extracted. Dashed orange lines indicate the values between which the linear fit was taken. b) Measured tunnel rates $\Gamma_{in/out}$ are plotted against the energy $E$, where $E=0$ represents the starting conditions of the experiment. The lines show fits to $\Gamma_{in} = \Gamma_0(1 + bE)f(E)$ and $\Gamma_{out} = 2\Gamma_0(1+bE)(1-f(E)$).}
    \label{fig:tunnelrates}
\end{figure}

Tunnel rates $\Gamma_{in/out}$ near the relevant charge transition were measured using the feedback technique developed in Ref.~\cite{Hofmann2016}, and described in detail for our setup in Ref.~\cite{BarkerFDR2022}. The rates were measured for two reasons. The first was to obtain the lever arm $\alpha$, which allows for the translation between the applied gate voltage $V_{g1}$ and the energy shift of the quantum dot level. The second was to use the tunnel rate to theoretically calculate the dissipated work with Eq.~(\ref{eq:wdiss}).\\
\\To extract $\alpha$, we note that the tunnel rates satisfy a detailed balance condition $\Gamma_{in}/\Gamma_{out} = 0.5\mathrm{exp}(-E/k_BT) = 0.5\mathrm{exp}(\alpha V_{g1}/k_BT)$. In Fig.~\ref{fig:tunnelrates}a), we plot $\mathrm{ln}(\Gamma_{in}/\Gamma_{out})$ against the gate voltage and fit a straight line to get $|\alpha| = 1.578\times10^4k_BT$/V. The fit was carried out between \SI{-0.15}{\milli\volt} and \SI{0.15}{\milli\volt} to avoid the effects of overestimation of the rates at low tunneling probabilities due to the finite measurement time of \SI{5}{\second}.\\
\\Figure~\ref{fig:tunnelrates}(b) shows the measured tunnel rates as a function of the quantum dot energy, together with fits to $\Gamma_{in} = \Gamma_0(1+bE)f(E)$ and $\Gamma_{out} = 2\Gamma_0(1+bE)[1-f(E)]$. From these fits, $\Gamma_0 = 59$~Hz and $b = -0.034/k_BT$ were obtained, giving an energy-dependent tunnel coupling $\Gamma = \Gamma_0(1+bE)$ used in Eq.~(\ref{eq:wdiss}).
\end{document}